\documentclass[twocolumn,secnumarabic,amssymb, nobibnotes, superscriptaddress, aps, pre]{revtex4-1}
\usepackage{amsmath}
\usepackage{graphicx}
\usepackage[dvipsnames]{xcolor}

\begin{document}

\title{Phase space consideration of low energy electron injection for Direct Laser Acceleration}%

\author{E. Starodubtseva}
\email{starodubtceva.em19@physics.msu.ru}
\affiliation{Faculty of Physics, Lomonosov Moscow State University, 119991, Moscow, Russia \looseness=-1}

\author{I. Tsymbalov}
\affiliation{Faculty of Physics, Lomonosov Moscow State University, 119991, Moscow, Russia \looseness=-1}
\affiliation{Institute for Nuclear Research of Russian Academy of Sciences, 117312, Moscow, Russia \looseness=-1}

\author{D. Gorlova}
\affiliation{Faculty of Physics, Lomonosov Moscow State University, 119991, Moscow, Russia \looseness=-1}
\affiliation{Institute for Nuclear Research of Russian Academy of Sciences, 117312, Moscow, Russia \looseness=-1}

\author{K. Ivanov}
\affiliation{Faculty of Physics, Lomonosov Moscow State University, 119991, Moscow, Russia \looseness=-1}
\affiliation{Lebedev Physical Institute of Russian Academy of Sciences, 119991, Moscow, Russia \looseness=-1}

\author{A. Savel'ev}
\affiliation{Faculty of Physics, Lomonosov Moscow State University, 119991, Moscow, Russia \looseness=-1}
\affiliation{Lebedev Physical Institute of Russian Academy of Sciences, 119991, Moscow, Russia \looseness=-1}

\date{February 2022}%

\begin{abstract}


Feasibility of ionization injection for Direct Laser Acceleration (DLA) of electrons up to hundreds of MeV was studied analytically. Criteria for effective injection determining range of background and in-channel plasma parameters, laser intensity, etc. were found using phase portraits of the system deduced from the simplified analytical model. The found optimal trajectory in the phase space corresponds to the electron with low (few eV) initial energy experiencing $\sim$10$^9$ times energy gain. For this to occur, electron density should be a few percent of the critical density, while the in-channel electron density should be $\sim$3 times lower. The analytically obtained dependence of the energy gain on the initial electron longitudinal and transverse momenta corresponds well to the results of exact numerical simulations of an electron motion in the plasma channel. Developed approach can form the basis for for further studies of electron injection in DLA varying plasma and laser parameters as well as initial electron energies.

\end{abstract}
\maketitle

\section{\label{sec:level1}Introduction}

Laser plasma acceleration of charged particles is one of the most promising areas of ultraintense lasers applications \cite{malka2008principles}. Today the most popular electron acceleration schemes are laser wakefield acceleration (LWFA \cite{doi:10.1063/1.37621}) and direct laser acceleration (DLA \cite{doi:10.1103/PhysRevLett.83.4772}). While the LWFA with 1 PW laser has been shown to produce high-quality quasi-monochromatic electron bunches with energies up to 8 GeV \cite{PhysRevLett.122.084801}, a divergence of 0.2 mrad and charge of several hundred pC, the DLA usually yields the quasi-exponential energy spectrum, but much higher charges (up to hundreds of nC \cite{doi:10.1088/1367-2630/abdf9a}). It is also can be implemented on tabletop femtosecond laser systems with 1-10 TW peak power \cite{doi:10.1088/1361-6587/ab1e1d, tsymbalov2020efficient}. 

One of the key issues for either acceleration scheme is the injection of the initial electrons \cite{doi:10.1063/1.3469581} since it strongly determines the parameters such as charge, energy, and divergence of the  electron beam obtained. The most established injection schemes to date are ionization injection \cite{doi:10.1088/0741-3335/58/3/034011} and wavebreaking of plasma waves \cite{doi:10.1103/PhysRevE.58.R5257}. Both in principle can be used for the DLA, however, ionization injection yields electrons with a low (few eV) initial energy, while wavebreaking -  subrelativistic ones. 

The optimum conditions for injection of high-energy electrons for the DLA are well discussed \cite{doi:10.1063/1.874154, doi:10.1063/1.873242, doi:10.1063/1.4964901, doi:10.1063/1.4975857, doi:10.1103/PhysRevE.98.033206,doi:10.1088/0741-3335/56/8/084006, doi:10.1103/PhysRevAccelBeams.24.041301, doi:10.1103/PhysRevLett.114.184801, doi:10.1088/1361-6587/aad76f, doi:10.1063/1.4766166, doi:10.1017/S0022377815000434, doi:10.1063/1.4951715}.
It has been noted that the energy gain obtained by electron depends on the phase of the laser field it has been injected into \cite{doi:10.1017/S0022377815000434,doi:10.1088/1361-6587/aad76f}. Rigorous theoretical analyses for the specific DLA models have shown that the energy gain depends specifically on the relative phase of the laser radiation and the Eugen electron's oscillation in the plasma channel \cite{doi:10.1063/1.874154,doi:10.1063/1.873242,doi:10.1063/1.4964901}.
Besides of the injection phase, the optimal initial electrons’ energies have been also analysed \cite{doi:10.1063/1.873242,doi:10.1063/1.4964901,doi:10.1063/1.4975857,doi:10.1103/PhysRevE.98.033206,doi:10.1088/0741-3335/56/8/084006,doi:10.1103/PhysRevLett.114.184801,doi:10.1088/1361-6587/aad76f,doi:10.1063/1.4766166,doi:10.1017/S0022377815000434,doi:10.1063/1.4951715}. Role of the electron transverse energy in the acceleration was demonstrated analytically \cite{doi:10.1063/1.873242} and the results \cite{doi:10.1063/1.4975857,doi:10.1088/0741-3335/56/8/084006,doi:10.1103/PhysRevLett.114.184801,doi:10.1088/1361-6587/aad76f} indicated that the optimal electron's injection condition corresponds to a quite large value of its transverse energy: $p_{0 \perp}/mc \approx (4 \div 6)$ ($m$ - electron mass, c - speed of light). Here and below $\perp$ and $\parallel$ notation indicate the direction perpendicular or parallel to laser radiation propagation direction, respectively. Several works have also demonstrated an importance of the electron initial longitudinal momentum for acceleration \cite{doi:10.1103/PhysRevLett.114.184801,doi:10.1063/1.4951715}. In \cite{doi:10.1063/1.4951715} it was highlighted that electrons are accelerated most efficiently when injected at small angels to the plasma channel axis: $p_{0 \perp}/p_{0 \parallel}\approx (0.15 \div 0.40)$.  All mentioned results demonstrate the possibility of an efficient energy gain by already pre-accelerated electrons.

However, there is not much discussion of the DLA of low-energy (i.e. created after ionization) electrons. To our knowledge, the only work in which this effect was discussed explicitly is \cite{doi:10.1063/1.4964901}. Therefore the goal of this paper was to understand the feasibility and optimal conditions for the ionization injection in DLA. 
We extensively analyzed the simplified DLA model for the case of electrons with low initial energies, reducing it to a qualitative analysis of the motion of an electron in a plasma channel. Further studies were conducted by analyzing resulting phase portraits. We also developed techniques to select a phase portrait topology corresponding to the acceleration of low-energy electrons at an extended range of initial phases. Therefore parameters at which ionization injection of electrons in DLA can be efficient were established. Developed approach in principle can be used to describe and optimize DLA electron acceleration for arbitrary initial and final energies.

\section{Motion of relativistic electron in the plasma channel}%

Let us first describe the basic interaction scenario and simplified analytical model of the DLA. When a high-intensity laser pulse interacts with the target and its power significantly exceeds the power of self-focusing, the formation of a plasma channel is possible \cite{doi:10.1063/1.871727}. The ponderomotive action of the laser pulse pushes electrons off the axis, thereby forming a radial electric field. Electrons that move along the direction of propagation of the laser pulse create an azimuthal magnetic field. Electrons moving in a combination of these fields under certain conditions experience energy gain. This is the essence of the DLA \cite{doi:10.1103/PhysRevLett.83.4772} mechanism.

Let us consider a symmetric infinitely long hydrogen plasma channel along z-axis with static radial electric and azimuthal magnetic fields:
\begin{equation}\label{e1}
   E^{s}_x=-2e\left( n_{e_{in}}-n_{e_{out}}\right) \pi x 
\end{equation}
\begin{equation}\label{e2}
   B^{s}_y=-2en_{e_{in}}\pi x 
\end{equation}
where $n_{e_out}$ - electron density outside the channel, which is equal to ion density in our model, $n_{e_in}$ - electron density inside the channel, $e$ is electron  charge. We will consider an electron moving in an $y=0$ plane.
It should be noted that in the model of hydrogen plasma we can neglect ionization by laser radiation.

Laser radiation is defined as an uniform plane wave:
\begin{equation}\label{e3}
   E^{L}_x=E_0^{L}\exp{\left(i\left(\omega t-kz\right)\right)}
\end{equation}
\begin{equation}\label{e4}
   B^{L}_y=\eta E_0^{L}\exp{\left(i\left(\omega t-kz\right)\right)}
\end{equation}
where $E_{0}^{\left( L\right) }=A\dfrac{\omega }{c}$ - laser electric field amplitude, $m$ is electron mass, $A$ - vector potential amplitude, 
$\eta$ - plasma refractive index ($\eta=\sqrt{1-n_{e_in}/n_{cr}}$).

\begin{figure}[b]
\includegraphics[width=1\linewidth]{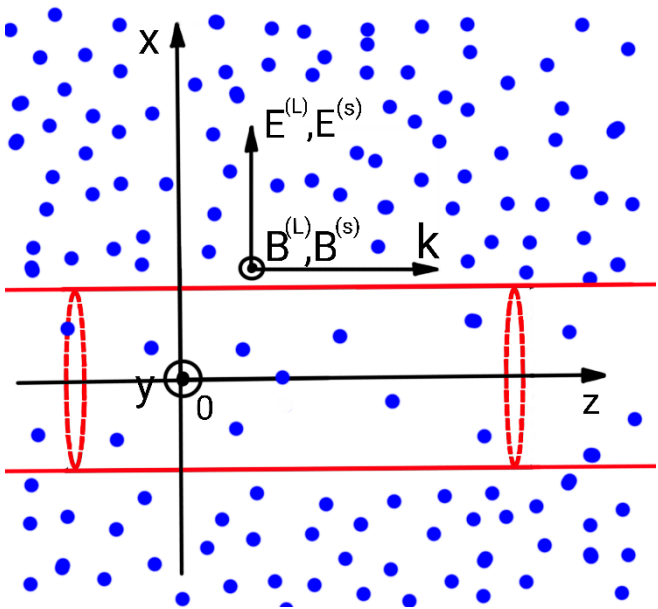}
\caption{\label{fig1} The schematic view of the plasma channel with electric and magnetic fields}
\end{figure}

This model uses a lot of approximations. The main ones are: (i)  the plane wave is a very crude description of a real laser pulse, and (ii) the plasma channel fields are in fact not static. However, it can be shown that such model can successfully predict maximum achievable electron energy gain in more realistic conditions \cite{doi:10.1063/1.874154}. Moreover, the indisputable advantage of this simple model is the possibility to study it analytically. Below we will demonstrate how by employing rather simple calculations one can estimate optimal initial parameters required for effective acceleration of electrons and get a deeper insight on the parameters that affect injection for the DLA. 

An electron motion in the combination of the static plasma channel fields and electromagnetic laser fields  \eqref{e1}-\eqref{e4} can be described as:

\begin{equation}\label{eq11}
\dfrac{dp_x}{dt}=-e\left[ E_{x}^{s}+E_{x}^{ L}-\dfrac{v_{z}}{c}\left( B_{y}^{s}+B_{y}^{L}\right) \right]
\end{equation}
\begin{equation}\label{eq2}
\dfrac{dp_{z}}{dt}=-e\dfrac{v_{x}}{c}\left(B_{y}^{s}+B_{y}^{L}\right)
\end{equation}
\begin{equation}\label{eq1}
mc^{2}\dfrac{d\gamma}{dt}=-e\left( \textbf E\cdot \textbf v\right) =-e v_{x}\left( E_{x}^{s}+E_{x}^{L}\right)
\end{equation}
where $\gamma=\sqrt{1+\dfrac{p_{x}^{2}+p_{z}^{2}}{m^{2}c^{2}}}$ - Lorentz factor. Electron trajectory will be that of anharmonic oscillator with Eugen frequency $\omega_b=\dfrac{\omega _{p}}{\sqrt{2\gamma}}$, which is named betatron frequency in the DLA case, and driving force frequency $\omega - kv_z$, where $\omega _{p}=\sqrt{\dfrac{4\pi e^{2}n_{i}}{m}}$ - plasma frequency.

Eqs. \eqref{eq11}\eqref{eq2}\eqref{eq1} have an integral of motion \cite{doi:10.1063/1.4964901}:
\begin{equation}\label{eq3}
I_{0} =\gamma +\dfrac{e^2\pi}{mc^{2}}\left( n_i-n_e(1-\dfrac{1}{\eta})\right)x^{2} -\dfrac{p_{z}}{\eta mc} 
\end{equation}

Note that the second term in \eqref{eq3} vanishes if $ x = 0 $, i.e on the plasma channel axis. Hereinafter we will assume that all the electrons start their motion with $ x(t=0) = 0 $.

\section{Injection and DLA  in the phase space}
Now we will use this model to study low-energy electron injection in DLA using phase portraits. 
Following \cite{doi:10.1063/1.874154} we applied the WKB approximation to simplify description of transverse motion of electrons using approximate expression:

\begin{equation}\label{eq12}
   x\approx \frac{v_{xA}(\gamma)}{\omega_b}sin\left(\int{\omega_b dt}\right), v_x\approx v_{xA}(\gamma)cos\left(\int{\omega_b dt}\right)
\end{equation}
where $v_{xA}$ - amplitude of an electron electron transverse velocity.

Using \eqref{eq12} and time-averaging over betatron oscillation period, equation \eqref{eq1} can be reduced to the system of two equations:
\begin{equation}\label{eqq4}
\dfrac{d\gamma }{dt}=-\dfrac{eE_{0}^{\left( L\right) }v_{xA}\left( \gamma \right) }{2mc^{2}}\cos \Phi
\end{equation}
\begin{equation}\label{eqq5}
\dfrac{d\Phi }{dt}=\omega -\omega _{b}-kv_{z}
\end{equation}
where $\Phi=\omega t-\int\omega_b d t-kz+\Phi_0$, $\Phi_0$ is an electron initial condition, which depends on initial phase of betatron oscillation ($\int\omega_b d t$) and initial phase of electromagnetic wave ($kz(t=0)$)  at which electron was injected. Equations \eqref{eqq4} and \eqref{eqq5} describe time evolution of an electron energy and the phase shift between the Doppler-shifted laser field and electron betatron oscillations.

\begin{figure}[b]
\includegraphics[width=0.49\linewidth]{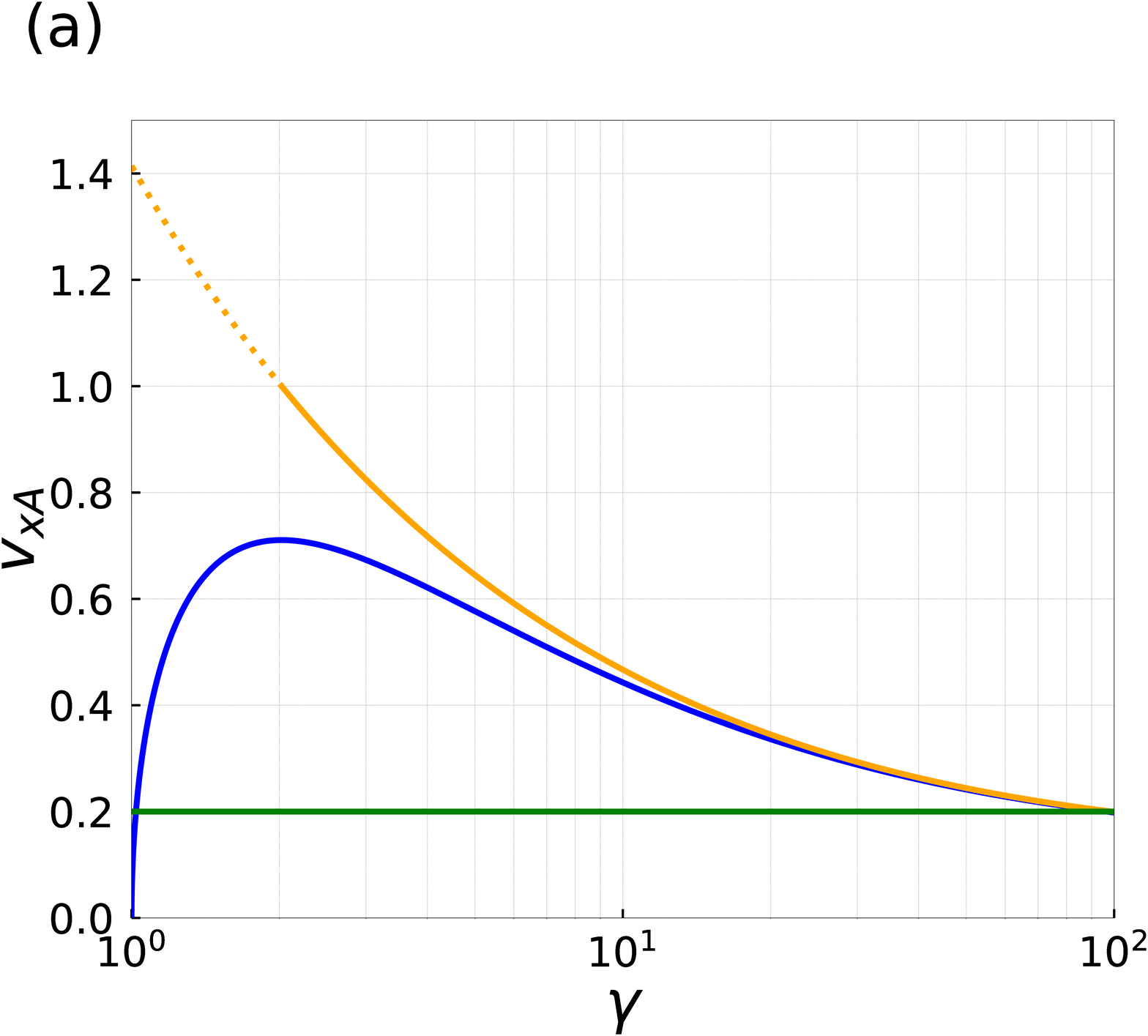}
\includegraphics[width=0.49\linewidth]{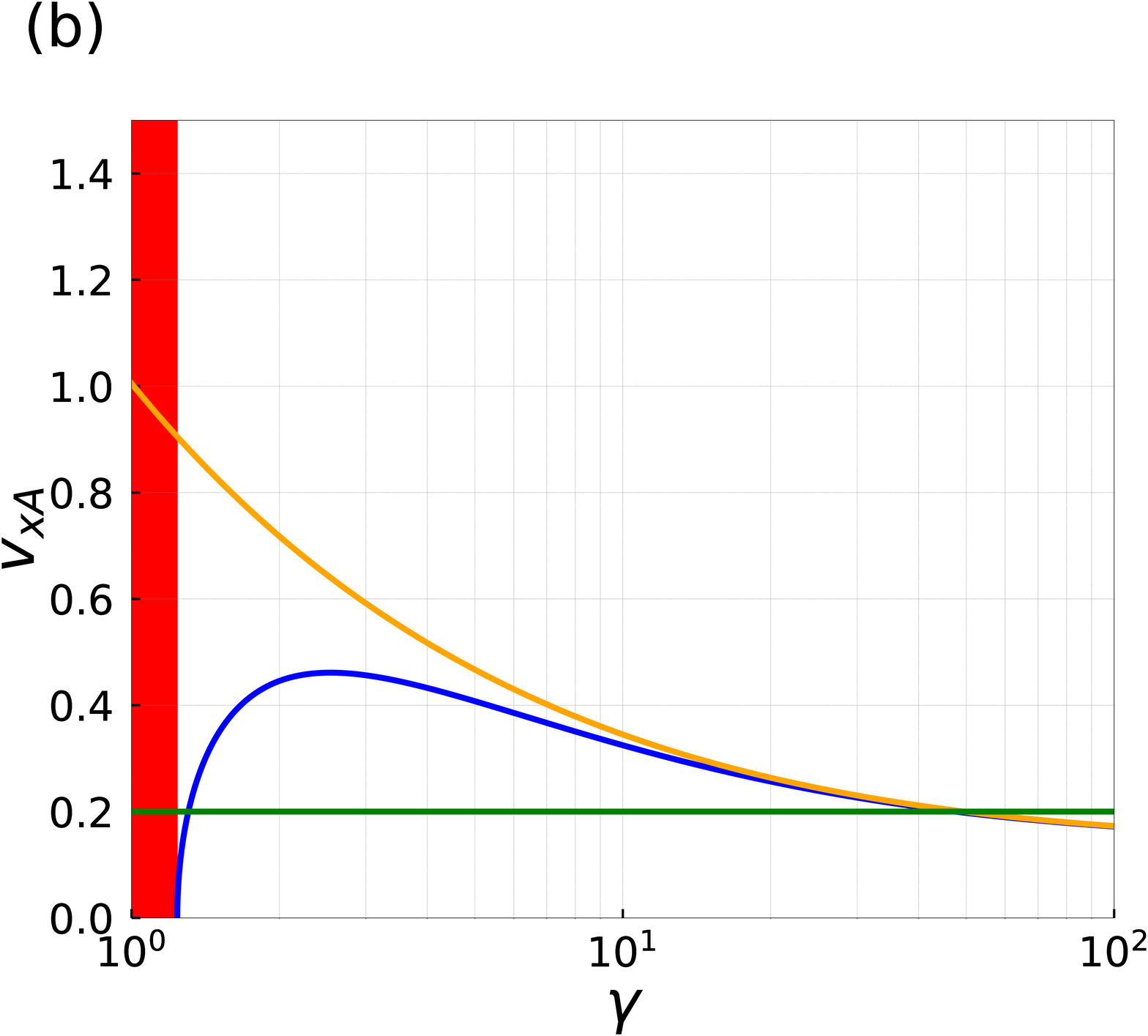}
\caption{\label{fig2}Dependence of the electron betatron oscillation velocity amplitude $v_{xA}$ on $\gamma$ calculated from \eqref{eq6} (blue), \eqref{eq7} (orange) and with $v_{xA}=0.2c$ (green)  \cite{doi:10.1063/1.874154} at  $\eta=0.99,  I_0=1.0$ (a), $\eta=0.99,  I_0=0.5$ (b). Red colored area in figure \ref{fig2}b indicates forbidden $\gamma$ values in \eqref{eq6}.}
\end{figure}

In  \cite{doi:10.1063/1.874154} $v_{xA}$ was considered to be constant. This is valid only for electrons with high initial energies ($\gamma \gtrsim 80$, see Figs.\ref{fig2}a,b). It is, however, not valid for electrons with low initial energies, which are the main focus of our work. In order to study the DLA of such an electrons one can express $v_{xA}$ from the integral of motion $I_0$ \eqref{eq3}:

\begin{equation}\label{eq6}
v_{xA}(\gamma)=c\sqrt{1-\dfrac{1}{\gamma ^{2}}-\eta ^{2}\left( 1-\dfrac{I_{0}}{\gamma }\right) ^{2}} 
\end{equation}

The resulting dependence $v_{xA}(\gamma)$ is shown in Fig.\ref{fig2} in blue. 
The system  \eqref{eqq4}--\eqref{eq6} can be solved numerically only. However, when electron transverse velocity $v_{xA}$ is much less than the speed of light, the paraxial approximation can be used \cite{doi:10.1063/1.4964901}:
\begin{equation}
v_{x}^{2}+v_{z}^{2}\approx c^{2}
\end{equation}
\begin{equation}
\dfrac{v_{z}}{c}\approx 1-\dfrac{v_{x}^{2}}{2c^{2}}
\end{equation}
\begin{equation}\label{eq7}
v_{x_{A}}\approx c\sqrt{2\left( 1-\eta \left( 1-\dfrac{I_{0}}{\gamma }\right) \right) }
\end{equation}
The dependence $v_{xA}(\gamma)$  \eqref{eq7} is shown as orange line in Fig.\ref{fig2}. One can see that the paraxial approximation can not be used for low-energy electrons, which are precisely the ones that we get from ionization injection. However, the difference between the approximate \eqref{eq7}  and exact \eqref{eq6} solutions is less than 10\% for $\gamma>5$. 
 Further, we will discuss that why this discrepancy can be neglected and hence equation \eqref{eq7} would still be applicable.

\section{Phase space investigation of DLA}

System of equations \eqref{eqq4},\eqref{eqq5},\eqref{eq7} can be solved analytically, with the solution being:
\begin{equation}
\sin \Phi =F\left( \gamma \right) -\alpha
\end{equation} where $F\left( \gamma\right)=-\frac{I_{0}\eta(\eta-2)\ln\left(\sqrt{\left(1-\eta\right)\gamma+I_{0}\eta}+\sqrt{(1-\eta)\gamma}\right)}{\sqrt{2}a_{0}\sqrt{1-\eta}}$
$+\frac{\left((\eta-1)(\eta+2)\sqrt{\gamma}+2\sqrt{2}\frac{\omega_{p}}{\omega}\right)\sqrt{\left(1-\eta\right)\gamma+I_{0}\eta}}{\sqrt{2}a_{0}(1-\eta)}$

The quantity $\alpha$ characterizes the trajectory in the phase space ($\gamma,\Phi$) for a given initial values of $\gamma_0,\Phi_0$ and  can be found as:
\begin{equation}\label{eq13}
\alpha=F\left(\gamma_0\right)-\sin \Phi_0
\end{equation}

\begin{figure*}
\includegraphics[width=1\linewidth]{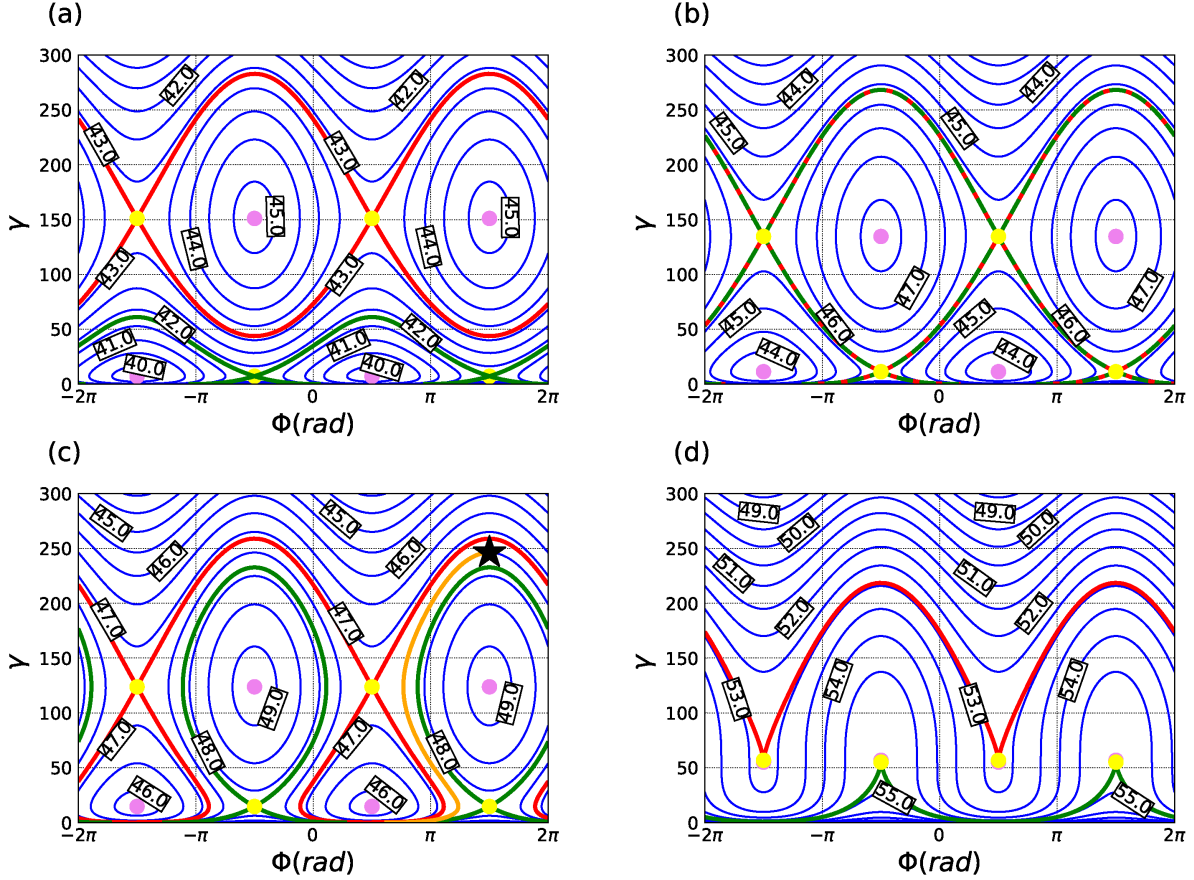}
\caption{\label{fig3}Phase portraits of the system \eqref{eqq4}\eqref{eqq5}\eqref{eq7} ($\eta=0.99$, $n_i=0.1n_{cr}$) for  $I_0=1.0$ (a),  $I_0=I_{0cr}=1.187$, where $I_{0cr}$ is $I_0$ value for homoclinic-heteroclinic bifurcation to occur (b),  $I_0=1.3$ (c),  $I_0=I_{0f}=1.706$, where $I_{0f}$ - $I_0$ value at which singular points merge (d).  Blue curves - phase trajectories for different values of $\alpha$ \eqref{eq13}, indicated as numbers. Red and green curves - separatrices, pink dots - center points, yellow dots - saddle points.  Orange curve in  Fig.\ref{fig3}c represents phase trajectory of interest for our study. Black asterisk marks  the maximum electron energy for this trajectory.}
\end{figure*}

It is convenient to use the phase space $(\gamma,\Phi)$ to study this system \eqref{eqq4}\eqref{eqq5}\eqref{eq7} (see Fig.\ref{fig3}). Each curve (phase trajectory) in Fig.\ref{fig3} represents evolution of the system for a specific initial conditions. Singular points (i.e. equilibrium states) of the system \eqref{eqq4}\eqref{eqq5}\eqref{eq7} can be found as:

\begin{equation}
\Phi _{cr_{1,2}}=\pm \dfrac{\pi }{2}+2\pi h
\end{equation}
\begin{equation}\label{eq20}
\gamma _{cr_{1,2}}=\dfrac{2\eta ^{4}I_{0}^{2}}{\left( \dfrac{\omega _{p}}{\omega }\mp \sqrt{\dfrac{\omega _{p}^{2}}{\omega ^{2}}-4\eta ^{2}I_{0}\left( 1-\dfrac{\eta }{2}-\dfrac{\eta ^{2}}{2}\right) }\right) ^{2}}
\end{equation}
where $h$ is an integer.
The points $\left( \Phi_{cr_2},\gamma _{cr_{1}}\right)$, $\left( \Phi_{cr_1},\gamma _{cr_{2}}\right)$ are centers (pink dots in Fig.\ref{fig3}). The points $\left( \Phi_{cr_1},\gamma _{cr_{1}}\right)$, $\left( \Phi_{cr_2},\gamma _{cr_{2}}\right)$ are saddles (yellow dots in Fig.\ref{fig3}). Separatrices (red and green curves in Fig.\ref{fig3}) pass through the saddle points and divide the phase-space into the regions, where the system behaves differently:
\begin{equation}
\sin \Phi -\sin \Phi _{cr_i}=F\left( \gamma \right) -F\left( \gamma _{cr_i}\right)
\end{equation}

For the considered system of equations, the topology of separatrices can be both heteroclinic (i.e. separatrix passes through each saddle once, see Fig.\ref{fig3}a) and homoclinic (i.e. separatrix passes through each saddle twice, see Fig.\ref{fig3}c), depending on the initial parameters. Figure \ref{fig3}b shows heteroclinic-homoclinic bifurcation of the phase portrait \cite{doi:10.1137/0521010}  (i.e. separatrix reconnection), through which these topologies switch. It takes place when separatrices pass through both saddles $\gamma_{cr_1}$ and $\gamma_{cr_2}$ and therefore merge. Hence one can find the $I_{0cr}$ value for this bifurcation to occur:
\begin{equation}\label{eq15}
2=F\left( \gamma _{cr_{1}}, I_0=I_{0cr}\right) -F\left( \gamma _{cr_{2}}, I_0=I_{0cr}\right)
\end{equation}

A further increase in $I_0$ leads to the merge (i.e. $\gamma _{cr_{1}}(I_0=I_{0f})=\gamma _{cr_{2}}(I_0=I_{0f})$) and disappearance of singular points and, correspondingly, separatrices (Fig\ref{fig3}d). It occurs at: 
\begin{equation}\label{eq16}
    I_{0f}=\dfrac{\omega _{p}^{2}}{4\omega ^{2}\eta ^{2}\left( 1-\dfrac{\eta}{2}-\dfrac{\eta ^{2}}{2}\right) }
\end{equation}

Electron acceleration regime of interest corresponds to Fig.\ref{fig3}c, as there exist phase trajectories (see, for example, the orange curve in Fig.\ref{fig3}c) connecting $\gamma \approx 1$ with large $\gamma$ values, with maximum attainable electron energy marked with a black asterisk. The phase portrait topology shown in Fig.\ref{fig3}a wouldn't be discussed further, as low-energy electrons are not captured there. Let us note, however, that it can still be used if one studies acceleration of electrons with larger initial energies ($\gamma >> 1$). The topology of the phase portrait after the merging of singular points (Fig.\ref{fig3}d) is poorly suited for acceleration, since electrons do not experience a substantial energy gain compared to other topologies.  It's important to remember that $I_{0cr}$ and $I_{0f}$ heavily depend on the plasma channel parameters (specifically $\eta$ and $n_i$).

\begin{figure}[b]
\includegraphics[width=1\linewidth]{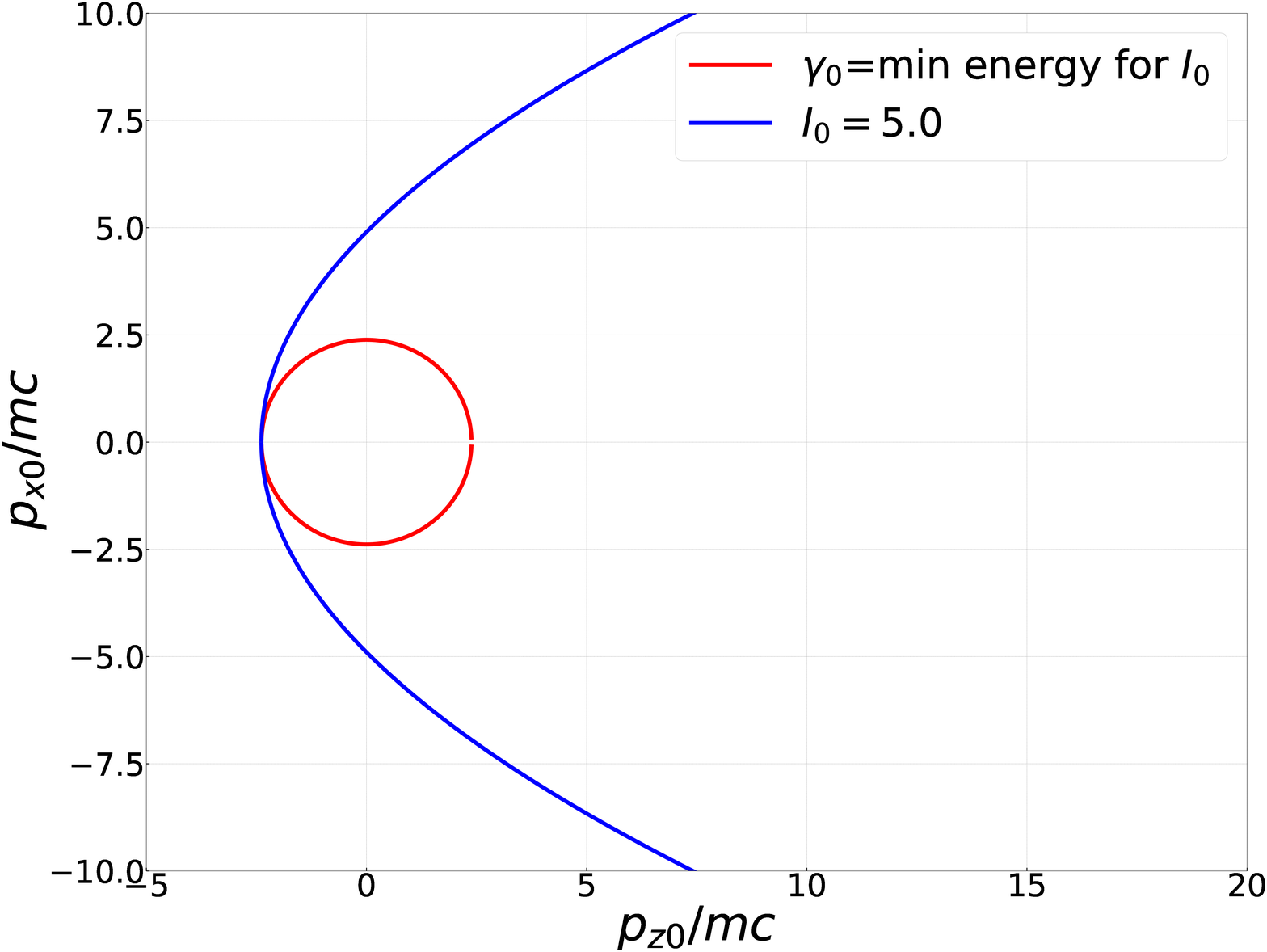}
\caption{\label{fig4} $I_0$ and $\gamma_0$ in the ($p_{x0}$,$p_{z0}$) coordinate plane. For every fixed $I_0$ (blue curve) $\gamma_0$ value can be greater than or equal to the red circle radius}
\end{figure}

We should also note that $I_0$ \eqref{eq3} and electron initial energy $\gamma_0$ are, in fact, dependent on each other. The easiest way to show this is to express them through initial transverse $p_{x0}$ and longitudinal $p_{z0}$ momenta: 

\begin{equation}
    I_{0}=\sqrt{1+\dfrac{p_{x0}^{2}+p_{z0}^{2}}{m^{2}c^{2}}}-\dfrac{p_{z0}}{mc\eta }
\end{equation}
\begin{equation}
    \gamma _{0}=\sqrt{1+\dfrac{p_{x0}^{2}+p_{z0}^{2}}{m^{2}c^{2}}}
\end{equation}

In coordinate plane ($p_{x0}$, $p_{z0}$) $I_0 = const$ is a hyperbola and $\gamma_0 = const$ is a circle (see Fig.\ref{fig4}). For a fixed $I_0$ only phase portraits with $\gamma_0 \geq \frac{-I_{0}+\sqrt{2 I_{0}^{2}-\eta^{2} I_{0}^{2}+1-\eta^{2}}}{1-\eta^{2}} $ would exist. As we are interested in acceleration of electrons with initial energies close to zero (i.e. $\gamma \approx 1$), only $I_0\approx1$ would satisfy that requirement. As was discussed earlier, $I_0$ should also correspond to the phase portrait topology shown in Fig.\ref{fig3}c, i.e. $I_{0cr} \leq I_0 \leq I_{0f}$. These two conditions can be rewritten as:
\begin{equation}\label{eq8}
    F(\gamma_{cr1},I_0=1)-F(\gamma_{cr2},I_0=1)<2
\end{equation}

\begin{equation}\label{eq14}
    \dfrac{\omega _{p}^{2}}{4\omega ^{2}\eta ^{2}\left( 1-\dfrac{\eta}{2}-\dfrac{\eta ^{2}}{2}\right) }>1
\end{equation}
where \eqref{eq8} reflects that separatrix reconnection \eqref{eq15} had already occured, and \eqref{eq14} reflects that singular points haven't merged yet \eqref{eq16}.

\section{Initial phases range}

One additional thing to consider is the range of initial phases $\Phi_0$, with which electrons can be accelerated. If this range is narrow it will result in poor injection efficiency even if conditions \eqref{eq8},\eqref{eq14} are satisfied. As $\Phi_0$ is determined, in part, by initial betatron oscillation phase, which is random for every electron, it is necessary to find acceleration regimes where electrons within the widest initial phase range would be captured.

\begin{figure}[h!]
\includegraphics[width=1\linewidth]{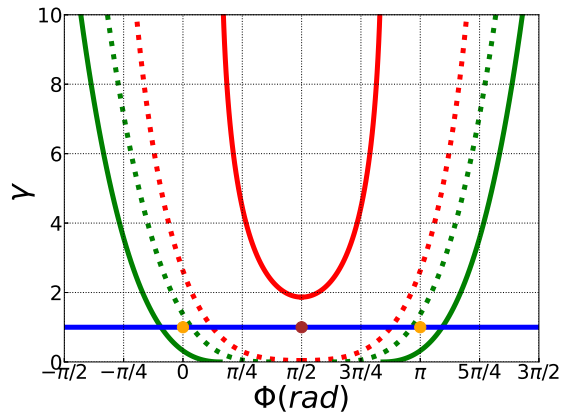}
\begin{flushleft}
\caption{\footnotesize The lower part of the separatrices corresponding to the topology of \ref{fig3}c (red and green). Initial phase ranges for injection energy $\gamma_in$ (blue line) is located between yellow points.  Solid curves indicate separatrices that satisfy \eqref{eq9},\eqref{eq10} ($\eta = 0.99$, $n_i=0.07n_{cr}$, initial phase range greater of equal to $0-\pi$), dotted curves - ones that do not satisfy \eqref{eq9},\eqref{eq10} ($\eta = 0.995$, $n_i=0.04n_{cr}$). Yellow and brown dots mark extreme values for meeting these conditions.}
\label{fig5}
\end{flushleft}
\end{figure}

Now we will find the conditions under which this range of initial phases will be greater than $(0-\pi)+2\pi h$ (for $\gamma \approx 1$), i.e. injection efficiency would be $>50\%$ for electrons with an initial energy $\gamma\approx 1$. Fig.\ref{fig5} provides an illustration for this conditions. Required phase range $(0-\pi)$ lies between the two yellow dots. To capture electrons  within this range of initial phases (i) the minimum of red separatrix has to be be larger than $\gamma=1$ (i.e. larger than $(\pi/2,1)$, brown dot) and (ii) the green separatrix has to to cross the $\gamma=1$ line at points $(0,1)$ and $(\pi,1)$:

\begin{equation}\label{eq9}
    F(1,I_0=1)-F(\gamma_{cr1},I_0=1)>0
\end{equation}
\begin{equation}\label{eq10}
    F(1,I_0=1)-F(\gamma_{cr2},I_0=1)<1
\end{equation}

\begin{figure}[b]
\includegraphics[width=1\linewidth]{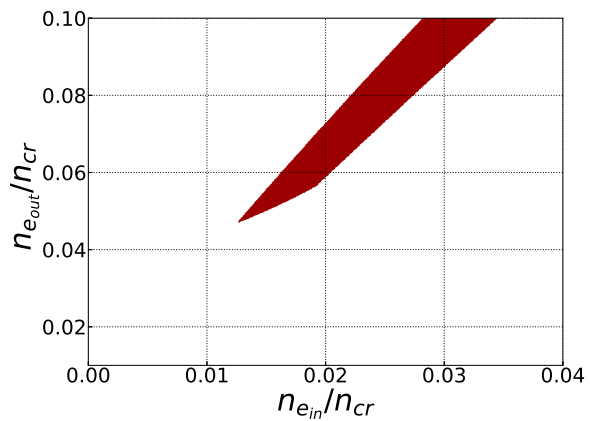} 
\caption{\label{fig6}The range of plasma channel parameters (shaded area) at which effective ionization injection takes place (conditions \eqref{eq8}\eqref{eq14}\eqref{eq9}\eqref{eq10}) are met)}
\end{figure} 

\begin{figure*}
\includegraphics[width=\linewidth]{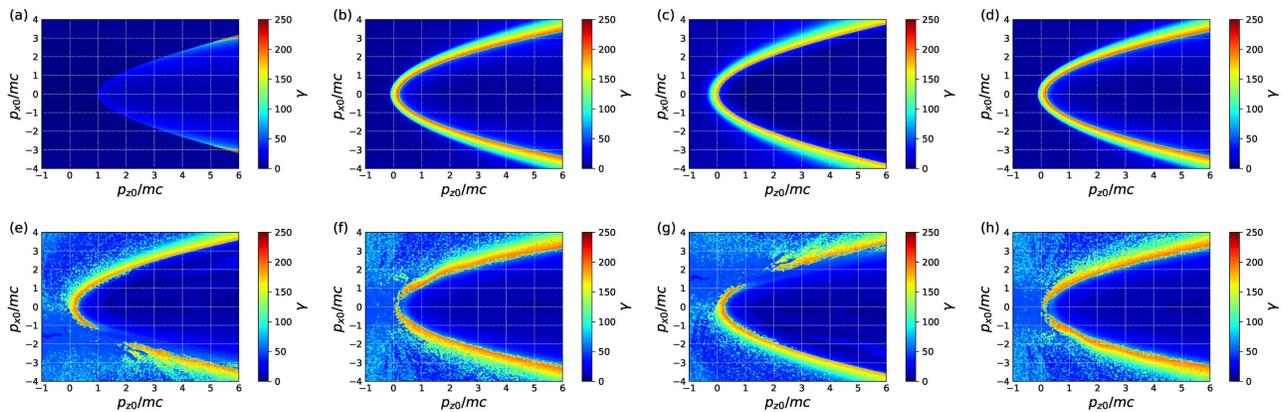} 
\caption{\label{fig7} (a-d): The analytically obtained maximum energy versus initial momenta $p_{x0}/mc$, $p_{z0}/mc$ for the fixed initial phase $\Phi_0$:  $\Phi_0=-0.5\pi$ (a),  $\Phi_0=0\pi$ (b),  $\Phi_0=0.5\pi$ (c), $\Phi_0=1.0\pi$ (d).
(e-h): Corresponding numerically obtained maximum energy versus initial momenta $p_{x0}/mc$, $p_{z0}/mc$ for the fixed electron initial position $z_0$:  $z_0=-\lambda/4$ (e),  $z_0=0$ (f),  $z_0=\lambda/4$ (g),  $z_0=\lambda/2$ (h). Plasma channel parameters for all figures are $\eta=0.99$ $n_{e_{out}}=0.07n_{cr}$.} 
\end{figure*} 

Thus, there is a set of channel parameters for which the above conditions are met and effective ionization injection might happen (see Fig.\ref{fig6}). Let us note, that these electron density values lie between that commonly used in both DLA and LWFA acceleration: tenths of $n_{cr}$ for the DLA  \cite{tsymbalov2020efficient} and thousandths of $n_{cr}$ for the LWFA. However, the required electron density can be experimentally achieved, for example, using high-density gas jet.

Figure \ref{fig7}a-d shows the analytically obtained dependence of the maximum achievable electron energy (corresponds to the highest point on the phase trajectory; see, for example, black asterisk in Fig.\ref{fig3}c) from its initial momenta $p_{x0}$, $p_{z0}$ under the \eqref{eq8}\eqref{eq14}\eqref{eq9}\eqref{eq10} conditions ($\eta=0.99, n_i=0.07n_{cr}$). In practice one can achieve this maximum energy by cutting the plasma channel at a correct length. It is also seen that  electrons with $p_{x0}\approx0$, $p_{z0}\approx0$ and initial phases $0 \leq \Phi_0 \leq \pi$ are indeed accelerated up to $\gamma\approx200$ energies for these parameters. 

The same dependence can also be obtained by numerically solving electron' equations of motion in a given fields of a plasma channel \eqref{eq11}\eqref{eq2}. Figure \ref{fig7}e-h shows such results (time range $0 - 2$ ps, $3*10^3$ steps). It can be argued that the maximum electron energies and corresponding initial momenta $p_{x0}(p_{z0})$, obtained from these two models, are in a good agreement.

What does not agree, however, is the initial phase range: no efficient acceleration takes place in Fig.\ref{fig7}a, but it's present in Fig.\ref{fig7}e. Firstly, this may be due to the fact that the initial phase $\Phi_0$ is determined both by the phase of the laser field and betatron oscillations, the latter one not specified in the numerical simulations. Secondly this discrepancy may be due to the stochastic processes, which are absent in the analytical model.

It should be noted that the model \eqref{eqq4}\eqref{eqq5}\eqref{eq7} (paraxial approximation) is correct despite its deviation at low energies from the more accurate model \eqref{eqq4}\eqref{eqq5}\eqref{eq6} (see Fig.\ref{fig2}, orange and blue curves respectively). This is due to the phase portrait topology being determined by the singular points of the system \eqref{eq15} \eqref{eq16}, with $\gamma_{{cr}_2} \approx 5$ for $I_0 \approx 1$ at all plasma channel parameters considered (see Fig.\ref{fig3}). For this $\gamma $ values two models match well. 
 
Another important difference between these models is that \eqref{eq6} (see Fig.\ref{fig2}, blue curve) in general has a range of forbidden $\gamma$ values (indicated in red in Fig.\ref{fig2}b). However for $I_0=1$, which is the value of interest, it vanishes, therefore only allowed values are present for the both models. 

Finally, note that an equivalent to \eqref{eqq4}\eqref{eqq5}\eqref{eq7} system of equations can be constructed for the case of circularly polarized laser radiation. All of the calculations still hold, with some changes in  coefficients (for example energy gain will be different as in \eqref{eqq4} $\dfrac{1}{2}$ will be changed to $\dfrac{1}{\sqrt{2}}$). 

\section{Conclusion}

We have studied in detail the analytically solvable model of electrons' direct laser acceleration (DLA). A 2D model with cylindrical hydrogen plasma channel given as static linear electric and magnetic fields and the laser given as a uniform plane wave was considered. 

An electron energy gain was investigated through phase portraits $ \gamma(\Phi)$ \eqref{eqq4} \eqref{eqq5}. The changes in the topology and the pattern of bifurcation of these phase portraits have been analyzed. The DLA regime in which low-energy electrons are efficiently captured and accelerated has been found. This regime is determined by a phase portrait topology, which, in turn, depends on the value of the motion integral $I_0$  \eqref{eq3}, defined by the initial parameters of electrons.

When certain conditions, i.e. parameters of the plasma channel (absolute refractive index and ion concentration) \eqref{eq8}\eqref{eq9}\eqref{eq10} are imposed to the phase portraits, electrons with approximately zero initial energy are accelerated effectively (up to $\gamma\approx200$ for $\eta=0.99$, $n_{e_{out}}=0.07n_{cr}$) within a wide range of initial phases.

These results are supported by the numerical integration of equations of motion. Analytical model and numerical integration results match both in the range of initial parameters and the maximum energies gained by electrons. However, conditions \eqref{eq8} \eqref{eq9} \eqref{eq10}, of course, do not exactly meet, therefore there is a discrepancy in the initial phase area.

To conclude, developed simplified analytically solvable model, supported by the numerical integration of equation of motion, demonstrates the possibility of ionization injection into the plasma channel at a rather wide range of its parameters and initial phases. 
Thus, this analytical model provides knowledge about the electron's motion in a plasma channel without numerical calculations and can form a basis for further study of electron injection in direct laser electron acceleration.

\begin{acknowledgments}
This work was supported by RSF Grant No. 21-79-10207. D.G. acknowledges the Foundation for Theoretical Research ‘Basis’ for financial support.

\end{acknowledgments}

\bibliographystyle{apsrev4-2}
\bibliography{references}

\end{document}